%% file: main.tex
 \documentclass[sigconf]{acmart}
\AtBeginDocument{%
  }

\usepackage[dvipsnames]{xcolor}
\usepackage{array}
\usepackage{booktabs}

\setcopyright{acmlicensed}
\copyrightyear{2027}
\acmYear{2027}
\acmDOI{XXXXXXX.XXXXXXX}
\acmConference[Conference acronym 'XX]{Make sure to enter the correct
  conference title from your rights confirmation email}{May 10--14,
  2027}{Pittsburgh, PA}
\acmISBN{978-1-4503-XXXX-X/2018/06}

\acmSubmissionID{3427}

\begin{document}

\title{Programming Prescience: Issues in Human Construction of Forecasting Bots}

\author{Priya Dhawka}
\email{pdhawka@uw.edu}
\orcid{0000-0002-7334-4950}
\affiliation{%
  \institution{University of Washington}
  \city{Seattle}
  \state{Washington}
  \country{USA}
}

\author{Benjamin Wilson}
\email{ben@metaculus.com}
\affiliation{%
  \institution{Metaculus}
  \city{Santa Monica}
  \state{California}
  \country{USA}
}
\author{Andrew Head}
\email{head@engineering.upenn.edu}
\orcid{0000-0002-1523-3347}
\affiliation{%
  \institution{University of Pennsylvania}
  \city{Philadelphia}
  \state{Pennsylvannia}
  \country{USA}
}

\renewcommand{\shortauthors}{Anon et al.}

\begin{abstract} 
AI models' growing capacity for complex reasoning has made them attractive tools for forecasting, or the systematic prediction of the future based on past and present data. However, should AI forecasting systems be adopted, human builders will shoulder the work of building and evaluating them. In this paper, we explore what this work looks like with a qualitative interview study with 15 builders of forecasting bots, sampled from among successful participants in a recent AI forecasting competition. Our analysis surfaces the challenges around (1) tuning a bot to acquire factual research, (2) implementing coherent reasoning, and (3) understanding how well one's bot actually does. 
During their work, builders frequently introduced complexity that  did not pay off, and tested in ways that were opportunistic and piecemeal. Our detailed account of their challenges helps clarify what problems should be targeted in upcoming innovations in process, tools, and human-AI teaming for forecasting.
\end{abstract}

\ccsdesc[500]{Human-centered computing~Empirical studies in HCI}


\keywords{Forecasting, Reasoning agents, Evaluation, Development}


\maketitle
\section{Introduction}
Forecasting is an essential method for predicting and communicating uncertain outcomes about the future. It is widely used to motivate decisions across myriad domains like election forecasting, public policy planning, business investments and financial trading among others~\cite{Armstrong_2001, Giangiulio_2026, Cai2024}. For example, within the US, major election cycles often include political polling and probabilistic election forecasts by expert human forecasters who make predictions about election outcomes to support public sensemaking around voting~\cite{Cai2024}. Human forecasting is also a costly and challenging effort involving stages of information gathering, reasoning about collected information and context, and producing a final forecast while reducing uncertainty at various stages~\cite{Tetlock_2005, Tetlock_Gardner_2015, Armstrong_2001}.

Recently, there have been indications that AI may become a viable tool to support forecasting. Halawi et al.~\cite{Halawi_Zhang_Chen_Steinhardt_2024} recently demonstrated language models achieving competitive performance with a human crowd on forecasting questions. Schoenegger et al.~\cite{Schoenegger_Park_Karger_Trott_Tetlock_2025a} observed that when people were given access to an AI during forecasting tasks, they achieved higher performance than when completing them alone. Models appear to be improving in their forecasting ability. The Forecasting Research Institute recently published ForecastBench as a benchmark for AI ability on the task~\cite{Karger_Bastani_Yueh-Han_Jacobs_Halawi_Zhang_Tetlock_2025}, and its public leaderboard posts continual improvements among the leading LLMs~\cite{ForecastBench}.

If AI is incorporated into forecasting, then the task of constructing and evaluating forecasting bots may become a significant human task. Aspects of this work are likely to diverge significantly from conventional human forecasting. For instance, a bot builder will have to figure out how to help a bot do background research over pools of sources that are more restricted to the bot than to themselves. Achieving good performance in forecasting may involve examining lengthy system outputs and translating them into ideas for productive changes.

The human-computer interaction research community has developed better approaches to predicting uncertain futures by refining and describing techniques for visualizing uncertainty~\cite{Hullman_Qiao_Correll_Kale_Kay_2019, kale_visual_2021, padilla_uncertain_2021, kay_when_2016, 10.1145/3290605.3300803, 10.1145/2858036.2858150, 10.1145/3613904.3642371} and providing tools for extrapolating and making decisions about the future with time series data~\cite{10.1145/2858036.2858150,10.1145/3313831.3376866, 10.1145/3491102.3502010}. We seek to complement this work with a deeper understanding of the task of building AI machinery that makes forecasts.

Recent community research has begun to describe this kind of work. For instance, one AI forecasting competition has made public surveys of bot builders' experiences. Among other things, it has revealed the effort involved in creating competitive bots (anywhere from hours to months) and identified features of bots that correlate with better performance (like use of web scraping)~\cite{spring2026survey}. We seek to provide researchers, tool-builders, and practitioners with greater color and depth around the human process of building forecasting AIs, and where the current ceilings are in this kind of work. What specifically are people trying to build? Where do they get their ideas? What bottlenecks are they hitting, both in regard to the underlying technology and their ability to make it do what they want? What workarounds are they devising?

To paint this picture, we conducted semi-structured interviews with 15 builders of forecasting bots. Builders were sampled from competitors in a recent AI forecasting tournament. They were all to some extent successful builders by nature of being prizewinners in the competition. They attempted a variety of implementation techniques, invested varying amounts of time and money, and ultimately achieved different levels of success. Our qualitative analysis provides a rich account of the current human work in building forecasting bots across the tasks of research, reasoning, and evaluation. Motivated by our observations, we discuss opportunities to improve process, tools, and human-AI coordination to improve AI-assisted forecasting.

\section{Related Work} \label{2}
We describe relevant concepts and prior research at the intersection of human forecasting, AI-assisted forecasting, general barriers to AI agent use, and the challenges of communicating uncertainty in predictions.

\subsection{Human Forecasting} \label{2.1}
Forecasting is the systematic process of making predictions informed by historical and present data. The forecasting literature describes best practices on how forecasters should produce reliable predictions~\cite{Tetlock_2005}. Notably, work from the Good Judgment Project identifies the reasoning practices that separate accurate human forecasters from the rest, such as decomposing questions, seeking contradicting evidence, and updating incrementally on new information~\cite{Tetlock_2005, Tetlock_Gardner_2015, Mellers_2015}, whereas the forecasting methodology literature more broadly catalogs principles around prioritizing simplicity in methods, combining forecasts, and structuring judgment for producing and evaluating reliable forecasts~\cite{Armstrong_2001}. Bot builders in our sample attempt to embed several of these best practices in their bots, albeit to varying levels of success.

\subsection{Forecasting with AI} \label{2.2}
Forecasting with AI comes with unique challenges that are understudied in current human-computer interaction and human-AI interaction research. Existing work by Schoenegger et al.~\cite{Schoenegger_Park_Karger_Trott_Tetlock_2025a} investigated the potential of using AI to augment human forecasts, finding that AI-augmented forecasts match and may even exceed performance by human forecasters. More recently, Paleka et al.~\cite{Paleka_Goel_Geiping_Tramèr_2025} identified two categories of pitfalls in evaluating AI forecasters relevant to our study: (1) users struggle to trust evaluation results due to temporal leakage, where models may access information published after an event's resolution date rather than genuinely predicting the outcome, and (2) users struggle to extrapolate from test results to real-world forecasting scenarios. Participants in our study also raised the temporal leakage concern, and we further identify the extrapolation problem as a limitation specific to tournament-style forecasting.

Prior work in this space also includes benchmarking resources for AI forecasting systems. Karger et al.~\cite{Karger_Bastani_Yueh-Han_Jacobs_Halawi_Zhang_Tetlock_2025} proposed a benchmarking dataset of forecast questions to evaluate the accuracy of AI-generated predictions while other benchmarks for AI-assisted, tournament-style forecasting relevant for our study include Halawi et al.'s work approaching human-level forecasting accuracy with language models~\cite{Halawi_Zhang_Chen_Steinhardt_2024}, ForecastQA's temporal event-forecasting question-answering challenge~\cite{jin2021forecastqa}, work examining whether language models can use forecasting strategies~\cite{pratt_2024}, Autocast++ that consists of retrieval-augmented approaches to the Autocast benchmark~\cite{yan2024autocast}, and the original Autocast dataset of real-world forecasting questions~\cite{Zou_2022}. Considering the nature of forecasting where there is a time delay between forecasts being generated and questions being resolved in real time, forecasters often require access to testing benchmarks that consist of historical questions that have already concluded as well as future questions that have yet to resolve. Thus, the main limitation of this existing benchmarking literature is that these datasets often have fixed date cutoffs, which is an issue that our participants also raised as forecasting questions often resolved after the training cutoff for most frontier models. 

A separate line of research also compares the performance of AI forecasters with human forecasters, with recent studies tracking AI's growing forecasting capabilities~\cite{Turtel_Franklin_Schoenegger_2025, Turtel_Wilczewski_Franklin_Skothiem_2026, Alur_2025, ForecastBench}. For instance, Schoenegger and Park~\cite{Schoenegger_Park_2023} and Abolghasemi et al.~\cite{Abolghasemi_Ganbold_Rotaru_2024} both evaluated large language models' predictive accuracy against human judgmental forecasting, while Schoenegger et al.~\cite{Schoenegger_Tuminauskaite_Park_Bastos_Tetlock_2024} compared the performance of an ensemble of LLMs against the aggregated accuracy of human forecasting crowds, finding that aggregated LLM predictions can at least rival the accuracy from human crowd forecasting tournaments. 
Additionally, ongoing research by Metaculus, a popular forecasting platform, has explored the problem of constructing an AI forecasting bot by surveying bot builders on their platform about the technical challenges of the process, documenting the forecasting strategies that resulted in the best performing bots. Metaculus' findings are publicly available as survey results and an accompanying guide for bot builders~\cite{Metaculus_AI_forecasting_2026}. In comparison, our study builds on this prior work by more closely investigating the strategies bot builders use to design successful forecasting bots and the workarounds they implemented to address the challenges they encountered in the process.


\subsection{Barriers to AI Agent Use} \label{2.3}
The literature on obstacles to AI agent adoption is large, spanning HCI, software engineering, and AI systems research, but we summarize a small number of papers within this space relevant to our study. At the systems level, Sapkota et al.~\cite{Sapkota_2026} proposed a broader systems view, distinguishing ``AI agents'' from more autonomous ``agentic AI'' and documenting recurring technical obstacles (hallucination, brittleness under distribution shift, coordination failures across tool calls, lack of persistent memory) alongside proposed solutions, some of which our participants also encountered, although in the context of forecasting where these challenges have unique impacts on the final forecast. Bansal et al.~\cite{Bansal_2024}, however, located the problem one level up, at human-agent communication rather than the underlying model. In particular, Bansal et al.~\cite{Bansal_2024} catalogued recurring breakdowns where an agent's actions, explanations, or requests for user input failed to align with what a user expects or needs, arguing that many failures attributed to model capability are in fact better understood as interaction-design failures. Separately, work by Shome et al.~\cite{Shome_Krishnan_Das_2026} surfaced a gap between industry aspirations for AI agents and the realities users experienced when actually deploying and relying on them, echoing challenges around understanding agent's behaviors, such as conservativeness during specific tasks and users expressing uncertainty about agentic behaviors, which we also observed among bot builders in our sample.

\subsection{AI Support for Communicating Uncertainty} \label{2.4}
Communicating uncertainty is a challenge in the forecasting process, where a forecast represents the probability of an event outcome, which also leaves it open to the decision-maker to decide how much confidence to place in and how to interpret the forecast. Prior HCI and visualization research has addressed the challenges and strategies of visually communicating uncertainty in forecasts, specifically highlighting the challenges of representing uncertainty in visualizations, that often result from the human inability to fully grasp the uncertainty associated with predictions~\cite{Hullman_Qiao_Correll_Kale_Kay_2019, kale_visual_2021, padilla_uncertain_2021, kay_when_2016}. This body of work largely centers around the representation of uncertainty, focusing on how uncertainty is visualized and how audiences interpret uncertainty visualizations
~\cite{10.1145/2858036.2858150,10.1145/3313831.3376866, 10.1145/3491102.3502010, 10.1145/3290605.3300803, 10.1145/2858036.2858150, 10.1145/3613904.3642371}. In our study, however, we found that participants raised other uncertainty concerns related to their bot's forecasts, including understanding the potential impacts of brittle changes and unnecessary complexity in their systems, and the absence of test datasets for questions among others.

\section{Methods} \label{3}
To understand the human issues in constructing AI forecasting bots, we conducted a qualitative interview study with 15 bot builders. We sought to answer the following research question in our study: 
\begin{itemize}
    \item RQ1: What challenges and bottlenecks do capable bot builders encounter when designing an AI forecasting bot?
\end{itemize}

\subsection{Research Setting} \label{3.1}
We recruited participants in collaboration with a platform, [Anonymous], that hosts regular human and AI forecasting tournaments for hobbyist and professional forecasters. With the help of one of the authors, A2 who also works at said platform, we recruited participants who had built successful AI forecasting systems for the last two tournaments during Fall 2025 and Spring 2026. In this AI forecasting competition held every quarter, [Anonymous], humans build AI forecasting bots, but given a question, their bots have to come up with the forecast autonomously. The forecasting platform, [Anonymous] offers monetary prizes, split from a shared pool of \$50k, where top performers in the tournament can make upwards of \$5k.

The forecasting platform provided a default bot template that included basic human forecasting principles and best practices that the human bot builders could customize at will. Customizations could incur at personal financial cost during the competition as participants needed to pay out of pocket for AI and other tools used outside of the small free usage tier provided by the forecasting platform. 

The AI forecasting tournament starts with a set of questions written by human experts, with specifications for which resolution criteria are included or excluded for each question outcome. In the competition, the bot is responsible for providing predictions for questions at a steady cadence. Questions were scheduled to resolve throughout the tournament. Submitted bots are given a question and a time limit to make a prediction, which is then compared against a resolution reflecting reality. Questions were of three kinds: 
\begin{itemize}
    \item Binary questions: simple yes/no forecasts where the bot submits a single probability of the event's occurrence.
    \item Numeric questions: the bot submits a full probability distribution over a continuous range of values instead of single point estimate.
    \item Multiple choice questions: the bot submits a probability distribution across a discrete set of mutually exclusive options, which sum up to 100\%.
\end{itemize}

Participants then build and test their AI systems (which we refer to as ``bots'' throughout this paper). At the end of each tournament, participants are ranked according to the accuracy of the forecasts produced by their AI bots, resulting in a leader board of bot builders, between whom a pool of prize money of \$50k, is distributed. Participant bots post their forecasts, with a rationale for their reasoning as a comment on the discussion threads for each question, making peer bot forecasts accessible to anyone participating in the forecasting tournament. The forecasting platform also runs shorter tournaments with a faster resolution turnaround every two weeks, that several participants mentioned using as an opportunity to test their bot performance. In terms of fostering engagement between participants, the platform offers a corresponding Discord channel for bot builders for tournament participants to discuss bot building and forecasting strategies.



\subsection{Participants}
To be eligible to participate in the study, participants were required to have been a recent prize winner in a recent AI forecasting tournament (prizes are awarded algorithmically based on the bot's rank and performance in coming up with timely and accurate predictions), be over 18 years old, and be willing to discuss their AI system. The second author identified prize winners in their most recent AI forecasting tournament (Fall 2025 and Spring 2026), to whom we then reached out via email to advertise the study. We chose to recruit prize winners as they were more likely to have successfully deployed a forecasting bot and implemented substantial forecasting strategies that (1) yielded positive outcomes and (2) may have required a substantial technical lift. 

Our final sample consisted of 15 participants, all identifying as male, within the age ranges of 20 to 60 years old, with the majority of our participants geographically located in the US and 6 participants residing outside of the US. Our study was approved by our university institutional review board. Participants had a variety of programming and forecasting expertise, which we summarize in \autoref{table:participants}, with the majority of participants working in professional software development positions ranging from interns to software engineers and sometimes AI/ML experts. In terms of coding expertise, 5 participants described having 6 months of programming experience whereas 2 participants were the most experienced builders in our sample with over 2 years of coding experience. Out of the 15 study participants, each participant managed a single bot set up for both forecasting tournaments, except for one pair of participants (P11a/b) who joined the tournament as a team, and who we interviewed as a team. 

\begin{table}[!ht]
    \centering
    \begin{tabular}{
        >{\raggedright\arraybackslash}p{1.5cm}
        >{\raggedright\arraybackslash}p{2cm}
        >{\raggedright\arraybackslash}p{2cm}
        >{\raggedright\arraybackslash}p{2cm}
    }
    \toprule
        Participant & Rank bucket & Prior Coding Experience & Prior Forecasting Experience \\
    \midrule
        P1 & Top 25 & No & Yes \\
        P2 & Top 15 & Yes & Yes \\
        P3 & Top 50 & Yes & Yes \\
        P4 & Top 5 & Yes & N/A \\
        P5 & Top 30 & No & None \\
        P6 & Top 10 & Yes & None \\
        P7 & Top 10 & Yes & None \\
        P8 & Top 15 & Yes & Yes \\
        P9 & Top 20 & Yes & No \\
        P10 & Top 20 & Yes & Yes \\
        P11a & Top 25 & Yes & Yes \\
        P11b & Top 25 & Yes & No \\
        P12 & Top 20 & Yes & Yes \\
        P13 & Top 5 & N/A & Yes \\
        P14 & Top 15 & Yes & Yes \\
    \bottomrule
    \end{tabular}
    \caption{A table showing the details of participants who chose to share additional information about their tournament rank, prior coding and forecasting experience. We show bucketed participant ranks ([1–5], [6–10], [11–20], etc) to obfuscate identity.}
    \label{table:participants}
    \Description{A table showing the details of participants who chose to share additional information about their tournament rank, professional backgrounds, and forecasting experience.}
\end{table}

\subsection{Procedure}
We used a semi-structured interview protocol with guiding questions to probe into our participants' process of building AI systems for forecasting. The first and senior authors, A1 and A3, conducted co-interviews with participants for up to an hour remotely over Zoom. We conducted all interviews as a pair except for two interviews. During the first half of the interview, we asked participants to walk us through their AI system's process of prediction for a single question from a recent forecasting tournament, asking them to describe the research and prediction set up of their forecasting bot and reasoning patterns or strategies they had successfully implemented or experimented with. During the second half of the interview, we probed participants for bottlenecks or challenges they experienced, including questions on how they tested their bots and reviewed output. We also probed participants for areas where they saw opportunities for future tools or resources that would help them improve their forecasting bot. We excluded from analysis any information that participants told us they did not want to share, such as private information related to their bot set up including anything they asked to keep out of the research manuscript. At the end of the interview, we offered participants a \$100 gift card as compensation for their time. Lastly, transcripts were generated by Zoom, which we then used a commercial LLM to fix errors before analysis. A sample semi-structured interview protocol is attached in ~\autoref{appendix:protocol}.

\subsection{Analysis}
We performed thematic analysis to extract themes describing related to participants' approaches for (1) implementing reasoning in their AI systems, (2) acquiring accurate and reliable research, (3) evaluating their system performance, and (4) tooling or research that could address the bottlenecks they encountered when building AI forecasting systems. The first author generated codes during the open coding stage until a saturation point was met, meeting regularly with the senior author to discuss coverage of relevant phenomena. Then they clustered them into higher-level themes based on their frequency in the data and similarities or differences in the transcripts. Our co-interview practice allowed us to debrief and reflect on participant strategies and challenges after the interviews~\cite{Velardo_Elliott_2021}, contributing to a richer understanding of participant experiences. Throughout the analysis and writing, we referred to a human-AI co-authored summary spreadsheet of the participants' detailed bot set ups and to the transcripts for fidelity and depth.

\input{findings_rev.2.tightn}

\section{Discussion}  \label{discussion}
\subsection{Takeaways}

Our observations of builders lead us to make the following provocations for how to improve the process, tooling, and human-AI teaming involved in forecasting.

\subsubsection{Process} For future builders of bots, we offer the following ideas:

\paragraph{Approach Complexity with Caution}
More complex approaches to designing AI forecasting bots do not necessarily correlate with capability or better performing bots in the tournament. Complex or conceptually intriguing reasoning approaches, including those inspired by real-world forecasting techniques, often did not pay off. Bots could succeed even when omitting patterns common to most others, as in the case of P2 whose bot made use of a single model without aggregation. Builders should not avoid complexity entirely---top performers like P4 and P13 integrated specialized research and fact checking components, for instance--but it may be appropriate to adopt a default skeptical stance towards changes one is considering making to a bot.

\paragraph{Opportunistic Review and Beyond}
We documented patterns of review that may be instructive to those getting started with bot building. Builders in our study reviewed bot outputs differently at three stages: (1) heavily and regularly during initial development (2) upon anomaly during deployment, and (3) systematically after a large number of predictions had been made. The first two stages were more common than the last. Builders largely did not review outputs after deployment unless an anomaly occurred. This does not necessarily mean that there is no value to be had in the remaining outputs. One builder, P8, described attempting to derive insight from bot outputs by processing them with AI. One opportunity for improved process might be to log and review predictions in ways that make it easier to learn how to optimize predictions that were not clearly bad but could nevertheless be improved.

Before questions were resolved, builders compared predictions to forecasts they made themselves and to community predictions. After resolution, builders compared their bot's reasoning to others'. Builders iterated on their bots by reviewing their performance on sample questions. Sample questions could be more or less useful based on their familiarity, alignment with their personal knowledge, diversity, half-life, and substance. We posit that future work might better establish how to build a maximally useful roster of questions under a fixed budget.

\paragraph{Testing and Iteration}
While builders mostly did not backtest their bots, we provide proof of possibility of doing so. P6 described a backtesting setup where they tested a reduced-complexity reasoning model against 30--100 questions using archived sources at an estimated cost of around \$1 per question. P8 described a related setup where they cached partial results from the pipeline, allowing for selective re-running of parts of the bot.

\subsubsection{Tooling}
Our analysis revealed several areas where builders could benefit from better tooling.

\paragraph{Intrepid Research Tools}
Some builders believed that better research methodology had the potential to greatly improve their bot. Two current bottlenecks in research was access to sources and the format of those sources. Builders extended access to additional information by querying external APIs and prediction markets and scraping web content. One builder created personal datasets that could be queried by their bot. These builders' bots could have benefited from deep research agents that went deeper and services that made specialized information discoverable and easily parseable.

\paragraph{Robustness to Bugs}
Small issues were capable of raising errors or steering a prediction in a wrong direction. P11 described a complication that issues in prediction can fly under the radar of standard error notification tools. Builders might benefit from better ways of detecting errors, both during deployment and before. One interesting opportunity may be to develop approaches to generating test questions that will maximally exercise the machinery of a bot pre-deployment and detect unusual output predictions.

\paragraph{Maximally Useful Questions}
Two participants called out Mantic AI's initiative to develop a test set of questions that can maximally differentiate the capabilities of different forecasters. P1 described wanting a similar discriminative set of questions for regular testing of their bot so that they could more rapidly converge on which parts of their bot would be the highest leverage to change.

\subsubsection{Human-AI Teaming}

Our study also helped to point out how the work of forecasting might be best organized within a human-AI team.

\paragraph{Research} Research was a perceived bottleneck for some of the participants. They described sources that were easy for a person to access and parse, but difficult for their bot. This helps to point out the role that people will likely have to play in accessing and reading sources in order to make the very best forecasts.

\paragraph{Making Sense} We speculate that there may be a significant role for people to play in sensemaking in various parts of the forecasting task. One participant described the difficulty of using a potentially useful piece of information---community predictions on prediction markets---because of differences in particulars of a prediction question between how it was posed to the bot and how it was written on the prediction market platform. We speculate that people might be able to make judgments about what kind of information might be derived and transferable from predictions on alternative but related questions. Furthermore, we also speculate that humans and AI may have to work together to make sense of suggestions that AI makes for how to improve a bot based on past performance. Understanding what to take action on likely requires taking into account a person's appetite for changing their bot and understanding of the kinds of situations in which it will be later deployed.

\subsection{Limitations \& Future Work}
This work has several limitations. Our sample skewed towards certain demographics, with the majority of our participants having prior coding experience, which potentially influenced the diversity of bot builder experiences in our study. Future work could investigate how the practices of users with less forecasting and coding experiences differ from our findings. We also acknowledge the limitations of using AI in the context of a forecasting competition, which include (1) questions are often written in a way for bots to try to come up with a prediction, and do not necessarily reflect real life forecasting conditions, (2) participants were not allowed to intervene in their bot's forecasting process due to the rules of the competition, which likely influenced the strategies they could implement. Future work could investigate the design of AI forecasting systems in real life, high-stakes contexts such as financial trading and election forecasting, where individuals and organizations are more likely to invest significant time and resources in building AI-assisted forecasting systems. 

Building on our participant insights, future work should also investigate how the structure of tournament-style forecasting and financial incentives influence the design of AI forecasting bots. For instance, future studies in this space should explore whether participant reasoning strategies generalize beyond competitions to other forecasting contexts like prediction markets, and the impacts and challenges of deploying competitive AI forecasting bots on live markets. Lastly, as AI-assisted forecasts become more common, we foresee opportunities to study how audiences and end users engage with these types of forecasts, particularly when considering downstream trust and accountability when using these types of forecasts in high-stakes settings where fact-checking is essential, such as in journalism or election forecasting.

\section{Conclusion}
We investigated the experiences of 15 bot builders and the challenges they encountered when building AI forecasting bots. Our analysis covers three tasks where bot builders hit ceilings in building their bots: research, reasoning, and evaluation. Among our observations are that complexity in AI forecasting bots often fails to translate into better forecasts, bot builders desire more resources to quickly and efficiently test their bots, and testing tended to occur opportunistically and piecemeal. Our work offers provocations to inform those entering bot building, those building tools to support it, and those looking to refine the human-AI relationship in forecasting.


\bibliographystyle{ACM-Reference-Format}
\bibliography{references}

\appendix \label{appendix}
\section{Semi-Structured Interview Protocol}\label{appendix:protocol}
\begin{enumerate}
    \item Can you walk me through your bot's process of prediction for a single question from a recent tournament? 
    \item \textbf{Follow-up as needed:} What does your bot do? 
    \item Can you show me any outputs or traces from the bot?
    \item Can you describe the reasoning patterns that you implemented in your bot? 
    \item Did it work? If not, what broke and how did you resolve these issues? 
    \item What other reasoning patterns would you implement if time and cost were not an issue? 
    \item \textbf{Optional:} What are some other approaches that you have tried and failed? 
    \item How well do you think your bot does? How do you know?
    \item What would you build if time and cost were not an issue? 
    \item \textbf{Optional if time permits} Tell me about a question that your bot has failed on in an interesting way.
    \item What were the bottlenecks to success that you noticed? 
    \item \textbf{For each bottleneck:} How did you get (or try to get) around this bottleneck? 
    \item How did you test your bot? 
    \item How did you iterate on your research pipeline? 
    \item What outputs or traces do you review? For example, do you review the final prediction, comments, Github action logs or other print outputs? 
    \item How much time do you spend manually reviewing your code, in one session or overall? 
    \item How do you design your logs or outputs to support testing and review?  What is your process for making sense of your logs? 
    \item \textbf{If not already covered} Did you learn or implement any strategies from other participants or the forecasting community? 
    \item How do you (and your bot) handle questions outside of your area of expertise? 
    \item \textbf{If not already covered:} What does your current tech stack lack? 
    \item \textbf{Optional if time permits:} Did you encounter other bottlenecks?
    \item What would make resolving these issues easier? 
    \item If there was one aspect of bot building or AI for forecasting that you would want researchers or tool-builders to make easier, what would it be? 
\end{enumerate}
\end{document}

%% file: findings_rev.2.tightn.tex
\newenvironment{spice}
  {\color{Plum}} 
  {}
\newcommand{\new}[1]{\textcolor{Plum}{#1}}
\newcommand{\iftime}[1]{\textcolor{gray}{If time: #1}}

\section{Findings} \label{5}

Our study focused on understanding the experiences of individuals building and deploying AI systems for forecasting. We discuss themes related to participants' experiences developing the research and reasoning abilities of their bots and evaluating their bots. Throughout, we call out factors that put a ceiling on the success of forecasting attempts.

\subsection{Research} \label{5.1}
Bots invariably began their predictions by looking for context on which to base them. The default template bot included set up instructions for the bot to gather research using the AskNews API\footnote{https://asknews.app/en} and to adopt the persona of a ``professional forecaster'' when coming up with a forecast. This template also uses ensembling or aggregating the results of multiple forecasts from various AI models into a final forecast, either as a mean or median of the multiple AI predictions. Participants could customize the template with their LLM of choice. The following challenges came up in getting them to do research well.

\subsubsection{Thoroughness}\label{thoroughness}


Quality predictions depended on quality research. As P4, builder of one of the best-performing bots, told us, ``the areas where you will beat the market by like 10 times are the ones where your research is really, really good.'' They described their bot as doing a ``whole load of research.'' P4 invested a great deal of effort in helping their bot find unusually useful information, exemplified by a particular question they described:

\begin{quote}
   ``My bot put 90\% on 29 cities\ldots{} First off, it loads up the website that it's about, and it requests an image of the website and analyzes that image\ldots{} it used that to manually count that there were 29 dots\ldots{} [and then it found that the page] was last written in 2022\ldots{} and it looked at the author. And then it had loaded up LinkedIn\ldots{} But it worked out he'd changed jobs from the organization that was recording this to a different job around when the blog stopped being updated, around when the last Freedom City was added. And it deduced from this that the chance of them updating it recently were very, very low.'' (P4)
\end{quote}

P4 told us that ``that question went very well'' for their bot. P4 described their approach as instilling ``curious agency'' in their bot, because ``if you can make an LLM that's curious about the things that matter, you can find stuff that others don't.'' To nudge their system to be curious, they prompted the bot to use the above example as a case study for future iterations. Several participants reported eagerness to push forward agentic approaches that granted more flexibility to the bot in seeking out background information (P4,P6, P13, P14). And several were implicitly working with agents, by invoking models that had deep research capabilities (P2, P4,P6, P11, P14). However, this flexibility came at a cost---P4 described adding switches to their bot so that it would adaptively shut off subsystems as it approached the per-question time budget, given that more flexible research had the potential to run for a long time.

Thoroughness involves not just accumulating the right background knowledge, but also checking one's work. P6 described a problem they perceived of models ``skipping cheap verification'' opportunities. P14 described their ``worst failure'' as involving a prediction that relied on a misreading of a foreign language website and how they recently added a fact checking module to the bot. When confronted with conflicting candidate predictions, P14's bot would use tools like web scrapers to search for ``facts that would sway the final forecast one way or the other.''



Like in P14's case, bots could rely on inappropriate information. P2 described their bot giving too much weight to overly biased sources. Broadly speaking, builders described the importance of seeking out a ``diversity'' of sources (P10 and P14), and many did, for instance by querying multiple sources (P3, P4, P6, P10, P13, P14) or implicitly querying specialized datasets (like social media posts) through calls to external models (like Grok) (P7, P10, P14). Builders also described conventional model issues affecting their research, including hallucinations (P7) and losing nuance from sources as their information is processed over multiple stages (P1, P3, P12).

\subsubsection{Access}\label{access}

Bots hit impasses in getting access to the best background information. As P3 put it,

\begin{quote}
``Whenever I went to look for the best data myself personally — even for the data that LLMs mentioned, `oh, it'd be nice if I had this' — they just didn't get it''
(P3).
\end{quote}



Why do LLMs fail to get the ``best data''? Builders described two reasons. First, sometimes the data is locked in a format that is inaccessible to the bot:


\begin{quote}
``a lot of data is\ldots{} often in, like,  an interactive dashboard that, you actually need to have the page load. You might even need to click around in a few different places to put the right filters in, etc. There is not necessarily a canonical URL that shows you exactly what the source that you need, and that data is often not indexed, so the search tools are not going to find it'' (P10).
\end{quote}

Second, access to the information could be restricted (P3 and P14). P3 provided the example of forecasting the price of plywood. They described the process of how one would seek out relevant information about the current cost, ruling out several options and arriving at Home Depot (an American hardware store) as a reasonable source of information. However, P3 shared how ``Home Depot's API is really hard to access these days---you have to go through a lot. And you don't want to just get banned because scraping their data is kind of illegal.'' P3 pointed out that in this situation one could actually get access to the Home Depot GraphQL, but because it is poorly-documented, it would require quite a bit of time for bots to research before it could be used to query useful information.


P14 described wanting ``tooling to pull any\ldots{} arbitrary piece of information'' that could provide ``robust'' and ``precise'' information about claims. In the absence of such tooling, builders took steps towards curating such a source by themselves. To overcome the issue of inaccessible data formats, they gave AIs control of the browser, for instance to take screenshots of pages that served reference material for forecasting questions (P10 and P13). Several builders programmed their bots to query APIs (P3, P6, P10). P3 built up their own knowledge base, reverse engineering data APIs and fetching their data into a local database that could be queried by their bot.

\subsubsection{Base Rates}\label{base_rates}


One practice of effective forecasters is to anchor predictions in initial best-guess base rates and update them as new evidence is encountered~\cite{Tetlock_2005, Tetlock_Gardner_2015}. For our participants, one of the roles of research is inform those base rates. Builders sometimes anchored their predictions in human predictions they found elsewhere because ``a lot of information is already kind of included in [them]'' (P13), presumably the information that others used justify placing a bet on the question. P13 explicitly encoded this approach in their bot:

\begin{quote}
“If we have a community prior, then I’m telling my bot that it should start from this. How strong is your prior? If it’s strong, then don’t let the next step influence you so much. Then use the news output from the assistants and update it depending on how relevant and trustworthy it is. Then state it again.” (P13).

\end{quote}

Builders sought out these base rates from multiple sources. For a time, they were able to anchor on the human ``community prediction'' from [Anonymous], and several did (P2, P13, P1, P11, P8) ; however, this was disabled, in part so that bot predictions would reflect situations with no available consensus~\cite{metaculus2026apichanges}. Some participants described anchoring on questions from prediction markets (P3, P6, P11, P13) such as Kalshi and Polymarket~\cite{kalshi2026, polymarket2026}. However, anchoring could be difficult to do well in this situation as ``you don't have the exact same question, the exact same resolution criteria, and so on'' for many of the questions posed (P13).

\subsection{Reasoning}
We interviewed builders when their bot forecasting performance was improving just by nature of improvement of the models they depended on (P3, P4, P6, P7, P8, P14). Here we focus not on that aspect---which was out of their control, but rather the design choices in their bots that they anticipated (though did not always) would improve their forecasts.


\subsubsection{Classic Forecasting Strategies}\label{forecasting_strategies}
Builders were often themselves forecasters, whether in a professional or amateur capacity (P1, P2, P3, P8, P10, P11a, P12, P13, P14). Many attempted to translate patterns of forecasting from their own work into the bots. They described trying to instill the right mindset in their bots, like using ``structured thinking,'' ``open-mindedness,'' using the ``wisdom of the crowd'' (P1), ``thinking hard,'' ``using math,'' (P2) ``superforecasters'' and ``Tetlock research''\footnote{Philip Tetlock leads a research lab that has popularized a system of forecasting described in ~\cite{Tetlock_Gardner_2015}. Superforecasters is a term coined to refer to people who demonstrate consistent ability in forming superior predictions.} (P10). Some of these goals were written explicitly into prompts. As described above, builders implemented techniques for finding reliable base rates (Section~\ref{base_rates}) and seeking out diverse sources (Section~\ref{thoroughness}). They experimented with patterns like incorporating Monte Carlo simulations (P4) and evaluating their bot's performance using Brier scores (P3).

Builders spoke at length about developing their bot to form predictions that aggregate over multiple individual predictions. This resembled the Delphi protocol from human forecasting~\cite{Tetlock_Gardner_2015, Armstrong_2001} (P8 named the method explicitly), where multiple forecasters make independent predictions, review each other's predictions, and submit revisions to be aggregated into a final prediction. Many bots had some form of aggregation approximating Delphi, as the default bot template had a basic form of the method built in. However, there was considerable variation---with participants aggregating over 2, 5, 8, and 54 predictions (P1), aggregating via medians or weighted averages (P3,P6, P7, P10), and aggregating over predictions from distinct models (P5, P6, P10, P14). Use of aggregation was common but seemingly not essential, with P2 achieving decent performance with a single model.

Some builders took effort to reduce undesirable interference between reasoning of various models used. An important aspect of procedures like Delphi is blindness of the individual predictors. P8 described enforcing blindness by architecting their bot so that individual predictions were not biased by each other. P7 described a variant in reasoning where a forecaster and critic evolved a prediction through iterative dialog. The critic was re-initialized on each turn to allow it to target different parts of the forecast, while the forecaster maintained context so that it could have a ``consistent sense of self'' and avoid vacillating in its forecasting approach.

\subsubsection{Optimizations}\label{optimizations}
Some of the work of building a bot involved local optimizations that more closely resembled patching over reasoning. Bots sometimes made obvious mistakes that were costly. P2 provided an example, where a question like ``will Google stock price go above X dollars?'' was predicted to be almost certainly true certain because the bot found it had previously been true, although the intent of the question was to ask about future price. P8 saw their model mistaking what the current date was at time of prediction leading to incorrect interpretation of prediction questions, where a date like ``May 25'' was seen as already having occurred in a previous year, rather than in the future. Builders described frequently fixing up these issues (P2). Others described adjusting the prediction distributions they output to the forecasting platform, calling tools to generate distributions (P3 and P6) or tweaking it, like P13, who added spikes to the distribution when the bot had high certainty in a specific prediction.

\subsubsection{Contending With Complexity}\label{complexity}
Some builders described very elaborate bots. P5, for instance, described an involved approach to aggregating over multiple predictions. Their bot aggregated over 8 predictions, each made with a confidence score. The final prediction was formed by using an average weighted by confidence score. However, P5 believed the aggregation suffered because models tended to produce confidence scores in a flattened range of 50--80. They therefore subtracted a constant from all scores until the minimum confidence was 15. Another example came from P1, whose bot produced output distributions on the basis of 54 individual predictions. 6 bots were prompted to predict 3 outcomes based on what they called ``low,'' ``mid,'' and ``high world'' scenarios, and then to refine those predictions into further low/mid/high variants ($6 \times 3 \times 3 = 54$). Their output distribution was formed by interpolating over those 54 predictions. These bot builders had invested considerable effort in their bot's reasoning strategies. P5 based their choices on what empirically seemed to work best; P1 had custom monitoring dashboards, and based their approach on ideas from when they had done forecasting professionally. However, these were among the less competitive bots in our sample, suggesting complexity does not always pay off.



P4, one of the better performers, provoked that ``complexity is unintuitively bad.'' They elaborated that ``generally I find that kind of you shouldn't overcomplicate your predictors. Every time I try and do something fancy, it's worse than just letting the base model do its thing.'' Participants experimented with many techniques, and many reported them not leading to substantial enough improvements to maintain. These techniques included tool calls to mathematical and statistical tools (P2), use of multiple predictions (P2), Monte Carlo simulations (P4), a post-prediction synthesis step (P4), prompt optimizations (P6), and use of probabilistic models like mixture models and-distributions (P6). Modest improvements could be less useful if accompanied by high token costs:

\begin{quote}
``I tried different like weighted averages where a review bot would weigh different forecasts\ldots{} and I tried also using different models\ldots{} it didn't have very good or I would say much improved results over the exploding token usage. So I think I used double or triple the tokens, but I didn't achieve scores double or triple'' (P12).
\end{quote}

This is not to say that the best bots are totally minimal. As described above, some of the best bots (such as P3, P13, and P14) involved instantiations of aggregation, fact-checking, agentic search, and other local optimizations. However, it does mean that many attempts to add useful techniques, including those grounded in participants' personal forecasting experience, do not substantially pan out. P11 conjectured that bots need a certain degree of ``flexibility'' to perform well; many of the adjustments described have the potential to limit this flexibility.

\subsection{Evaluation} \label{5.4}
Builders needed input on how well their bot predicted, but getting this input remains a tricky problem.



\subsubsection{Brittleness and Bugs}
Bots could fail in substantial ways that were not easy to predict. P2 and P8 described two examples that were triggered by only specific formats of questions that we covered in Section~\ref{optimizations}. Nor were they necessarily easy to detect. P8 described an additional example of a major issue that was not readily obvious: 

\begin{quote}
``there's this weird thing about how I'm doing some of the interpolation, where sometimes you accidentally do it in a slightly wrong way, and\ldots{} it just, like, destroys the template and just dumps a different thing there that shouldn't be there\ldots{} [the LLMs] still will give you an answer that probably is much worse than it otherwise would be, but\ldots{} it's not worse in a way that immediately is apparent that something is broken'' (P8).
\end{quote}

Furthermore, changes made to fix one problem could introduce others. P8 described a case where they made changes to a prompt for deriving consensus from their predictors that ``caused it [to] try do a net new forecast'' without the research data and ``was much worse than the individual forecasts that were feeding into it.'' Given the potential for a change to degrade unanticipated predictions, P14 shared that ``you need a lot of data to be able to justify doing some kind of adjustment, or you'll cause more harm than it was worth in the first place.''






\subsubsection{Testing}
Builders described at times how they would have found it useful to get more feedback on how well their bot was doing, especially after making changes to it (P6 and P8). One setup involved running the bot on a set of historical questions and computing the accuracy of its predictions, an approach called backtesting. P6 described their approach:

\begin{quote}
``I log my research and go back and feed it to typically a slightly smaller version of my ensemble to save money, and actually compute the peer score that we're evaluated on in the tournament, and use that for evals. But again, that's really noisy. I do compute statistics, of course, to try to be rigorous. And it typically requires at least 30 or 50, maybe 80 or 100 plus questions'' (P6).
\end{quote}

However, this approach was not common. The only other builder who described backtesting was P3, though they did not share the particulars of their approach. P8 designed their bot to cache research so that they could selectively evaluate their research stage, but did not integrate it into a backtesting flow. Backtesting is difficult because when a bot is tested on historical questions it is often ``taking in information from the future'' in its predictions (P13), finding clues or answers to how an event resolved from published research. In this way, a bot can outperform its actual ability for forecasting future questions. One potential solution involved restricting a bot to only use historical sources, though P3 described that information could still be leaked:

\begin{quote}
``Retrodiction/backtesting for forecasting is notoriously leaky\ldots{} you look into the actual logs [and see that] there's a little bit of leakage from---because they changed the article, and our SERP or search provider didn't really recognize that this had been changed, so the time-archived copy actually leaked information a little bit, just enough so the LLM didn't explicitly use it, but was tainted anyways'' (P3).
\end{quote}

Complicating the picture, testing is costly. If a testing suite involves 100 questions and a typical question costs on the order of \$1, as P6's did, ``if I want to do rigorous evals, call it \$100.'' Other bots cost more (P13 estimated 3 euros per question). For these builders, their bot would net a few thousand dollars in the best case if they performed well in the competition. Tests of P6's scale could only be run sparingly if a bot was to break even. P4 told us they would need ``way more money'' to do more backtests. P5 wished for historical source datasets they could use to backtest.

\paragraph{Monitoring}
Some participants made custom dashboards for monitoring their bots (P1, P3, P14). P2 and P11 set up their bot to send them email alerts on failure. However, these alerts were not silver bullets for catching failures: P11 described that sometimes ``it doesn't report an error, but it [also] doesn't predict.''


\subsubsection{Inspection}
The most common approach to evaluating one's bot was to review its outputs---its predictions, summaries of rationale, and other logs produced during execution. All participants reviewed their bots during except for P1, P5, P9, and P11a/b who reviewed after prediction. Review would often spike early during the bots development(P5, P10, P11a/b, P13). P5 described when developing the bot that initially ``I just sat there, I just, like, I'd run it, look at [the output], run it, look at it, run it, like, all day.'' The amount of time spent looking at the outputs could be significant. P11a/b estimated spending ``maybe half, or even over half'' of their total development time looking at logs, where most of that was spent on reviewing forecasting outputs. 

Later, when the bot was implemented and deployed, time reviewing logs decreased. Builders described spending ten minutes daily (P3), a couple of minutes for a handful of questions (P6) or a couple hours a month (P4), or even less time(P10 and P13) reviewing them. Review was often triggered by anomalies:

\begin{quote}
``I’m not checking every single answer. I’m mostly checking the ones where my bot says something very different from the community prediction. And then I check my bot's answers, and think about “Is my bot being very smart or very dumb?'' (P13)

\end{quote}

P2 similarly described looking at reasoning output only when they saw a ``crazy prediction.''

Builders could also review outputs in a batch during retrospectives. P8 analyzed the logs from their bot when a competition season concluded. This approach allows review of predictions across many questions when the resolution is known. P8 used AI to summarize strengths and opportunities for improvement for their bot based on the logs.

While reviewing outputs was never described as a significant bottleneck, it also did not seem trivial. P8 shared that ``knowing that something's awry is quite challenging.'' P5 described that ``it's really hard to, like, read through pages and pages of, you know, AI slop to determine some sort of signal.'' P10 put it that ``some of them might just be relatively long, and it just takes a little bit more time and and mental overhead'' to review them. Logs could indeed be quite long---P4 described generating 10MB of logs per prediction.

Review of outputs was described as involving ``intuition'' (P4) or was ``vibes-based'' (P10). Those who were experienced forecasters would sometimes perform forecasts of their own and compare the bots' output to them (P10 and P11). Other times, builders compared their logs to those of competitors. This could help them determine if a difference in their prediction arose from a ``reasonable deviation'' such as finding a piece of information other bots did not (P2). P6 and P8 described reviewing comments from other bots, with P6 drawing inspiration from competitor bot outputs. P13 described an intriguing variant where they attempted to anticipate the effect of a change to their bot by downloading earlier predictions and asking what would have been different if their change was in place.

To look for opportunities to improve, builders sometimes consulted reference prediction values. P3 described comparing their bot's predictions to live market predictions on Kalshi~\cite{kalshi2026}. Some of the bot builders had previously compared to aggregated community predictions made by human participants on [Anonymous] (P6).\footnote{This feature was since turned off by [Anonymous], in part to promote evaluation in ``the more realistic condition of no available consensus.''} Builders observed that human predictions may become less valuable as models continue to improve (P6). As P14 put it, such scores could be ``limiting, unless you really don't trust that you're gonna exceed community prediction ever.''
P1 pointed out how AI-generated forecasts impacted community predictions which were no longer reliable for testing: 
\begin{quote}
    `So I think it's mostly that if you ask the same LLM the same question or a similar question in a different way, the responses are more likely to be similar than if you're asking a different LLM'' (P1).
\end{quote}

\subsubsection{Samples of Questions}
When participants evaluated their bots, they needed to do so on specific questions. As described above, one strategy was to review questions from the competition that led to anomalous outputs. While developing and evolving their bots, however, builders needed to select sets of questions to test on.

Builders described selecting questions on the basis of:
\begin{itemize}
    \item \emph{familiarity}. For instance, P9 tested their bot on random factual questions to which they could find or easily derive an answer, like who is the president of a given country, music trivia, or basic equations and formulas 
    \item \emph{personal insight}. P6 described selecting questions ``where I have a strong opinion, and where I've seen bots---at least based on the community benchmark---get them wrong in the past.''
    \item \emph{diversity}. P5 and P11 described selecting samples that covered a diversity of question types and subject matter, perhaps because their bots would be evaluated against a diversity of questions.
    \item \emph{half-life}. P8 described testing against a recommended practice question of whether there will be human extinction by 2100. The benefit of this question was that ``the news changes\ldots{} but not that much to actually impact that most of the time. So it's relatively steady, constant thing to run predictions against.''
    \item \emph{substance}. P6 chose questions that involved some ``proactive thought\ldots{} connecting the dots, thinking beyond the exact single chain of thought and not drawing any conclusions.''
\end{itemize}


Some participants evaluated their bot's performance by entering it in a small-stakes biweekly tournament also offered by [Anonymous] (P7, P8, P14). For one participant, P1, this smaller tournament provided an opportunity to clarify what makes a useful question, which this smaller tournament did not seem to reliably provide \footnote{At the time of writing, question generation for this smaller, biweekly tournament changed from template questions to AI-generated questions and may have resulted in more substantial questions for testing.}. The questions either ``resolve very quickly,'' which presumably means they require less ambitious reasoning; or they have ``high random content,'' which presumably means that changes to a bot are unlikely to lead to reliably better performance on them. P1 and P10 both pointed out the value of having more useful questions. They pointed to an effort by Mantic AI\footnote{Mantic AI, \url{https://www.mantic.com/}, a private company building forecasting AI that also hosts forecasting tournaments for AI and humans.} to collect more useful forecasting questions, which were defined as those that maximally differentiate between forecasters.

Regardless of how questions were selected, one issue is that a small sample of questions could only be so informative. As P1 told us, ``it's easy to over-interpret the results of your questions because you have a relatively few questions of each type.'' P5, who evaluated against a set of 17 questions meant to cover different question types, described results of their tests as necessarily ``incredibly noisy.'' Even P6 described test results as being ``really noisy'' when they ran their backtest benchmark of 30--100 questions.

%% file: references.bib
@article{Hullman_Qiao_Correll_Kale_Kay_2019, title={In Pursuit of Error: A Survey of Uncertainty Visualization Evaluation}, volume={25}, ISSN={1941-0506}, url={https://ieeexplore.ieee.org/document/8457476/}, DOI={10.1109/TVCG.2018.2864889}, abstractNote={Understanding and accounting for uncertainty is critical to effectively reasoning about visualized data. However, evaluating the impact of an uncertainty visualization is complex due to the difficulties that people have interpreting uncertainty and the challenge of defining correct behavior with uncertainty information. Currently, evaluators of uncertainty visualization must rely on general purpose visualization evaluation frameworks which can be ill-equipped to provide guidance with the unique difficulties of assessing judgments under uncertainty. To help evaluators navigate these complexities, we present a taxonomy for characterizing decisions made in designing an evaluation of an uncertainty visualization. Our taxonomy differentiates six levels of decisions that comprise an uncertainty visualization evaluation: the behavioral targets of the study, expected effects from an uncertainty visualization, evaluation goals, measures, elicitation techniques, and analysis approaches. Applying our taxonomy to 86 user studies of uncertainty visualizations, we find that existing evaluation practice, particularly in visualization research, focuses on Performance and Satisfaction-based measures that assume more predictable and statistically-driven judgment behavior than is suggested by research on human judgment and decision making. We reflect on common themes in evaluation practice concerning the interpretation and semantics of uncertainty, the use of confidence reporting, and a bias toward evaluating performance as accuracy rather than decision quality. We conclude with a concrete set of recommendations for evaluators designed to reduce the mismatch between the conceptualization of uncertainty in visualization versus other fields.}, number={1}, journal={IEEE Transactions on Visualization and Computer Graphics}, author={Hullman, Jessica and Qiao, Xiaoli and Correll, Michael and Kale, Alex and Kay, Matthew}, year={2019}, month=jan, pages={903–913} }

@article{Karger_Bastani_Yueh-Han_Jacobs_Halawi_Zhang_Tetlock_2025, title={ForecastBench: A Dynamic Benchmark of AI Forecasting Capabilities}, volume={2025}, url={https://proceedings.iclr.cc/paper_files/paper/2025/hash/ea74e45a229dac70b5b63b28d8934db6-Abstract-Conference.html}, journal={International Conference on Learning Representations}, author={Karger, Ezra and Bastani, Houtan and Yueh-Han, Chen and Jacobs, Zachary and Halawi, Danny and Zhang, Fred and Tetlock, Philip E.}, year={2025}, month=may, pages={93943–93980}, language={en} }

@article{Mellers_2015, address={US}, title={The psychology of intelligence analysis: Drivers of prediction accuracy in world politics}, volume={21}, ISSN={1939-2192}, DOI={10.1037/xap0000040}, abstractNote={This article extends psychological methods and concepts into a domain that is as profoundly consequential as it is poorly understood: intelligence analysis. We report findings from a geopolitical forecasting tournament that assessed the accuracy of more than 150,000 forecasts of 743 participants on 199 events occurring over 2 years. Participants were above average in intelligence and political knowledge relative to the general population. Individual differences in performance emerged, and forecasting skills were surprisingly consistent over time. Key predictors were (a) dispositional variables of cognitive ability, political knowledge, and open-mindedness; (b) situational variables of training in probabilistic reasoning and participation in collaborative teams that shared information and discussed rationales (Mellers, Ungar, et al., 2014); and (c) behavioral variables of deliberation time and frequency of belief updating. We developed a profile of the best forecasters; they were better at inductive reasoning, pattern detection, cognitive flexibility, and open-mindedness. They had greater understanding of geopolitics, training in probabilistic reasoning, and opportunities to succeed in cognitively enriched team environments. Last but not least, they viewed forecasting as a skill that required deliberate practice, sustained effort, and constant monitoring of current affairs. (PsycInfo Database Record (c) 2025 APA, all rights reserved)}, number={1}, journal={Journal of Experimental Psychology: Applied}, publisher={American Psychological Association}, author={Mellers, Barbara and Stone, Eric and Atanasov, Pavel and Rohrbaugh, Nick and Metz, S. Emlen and Ungar, Lyle and Bishop, Michael M. and Horowitz, Michael and Merkle, Ed and Tetlock, Philip}, year={2015}, pages={1–14} }

@article{Schoenegger_Park_Karger_Trott_Tetlock_2025a, title={AI-Augmented Predictions: LLM Assistants Improve Human Forecasting Accuracy}, volume={15}, ISSN={2160-6455}, url={https://dl.acm.org/doi/10.1145/3707649}, DOI={10.1145/3707649}, abstractNote={Large language models (LLMs) match and sometimes exceed human performance in many domains. This study explores the potential of LLMs to augment human judgment in a forecasting task. We evaluate the effect on human forecasters of two LLM assistants: one designed to provide high-quality (“superforecasting”) advice, and the other designed to be overconfident and base-rate neglecting, thus providing noisy forecasting advice. We compare participants using these assistants to a control group that received a less advanced model that did not provide numerical predictions or engage in explicit discussion of predictions. Participants (N  (=)  991) answered a set of six forecasting questions and had the option to consult their assigned LLM assistant throughout. Our preregistered analyses show that interacting with each of our frontier LLM assistants significantly enhances prediction accuracy by between 24% and 28% compared to the control group. Exploratory analyses showed a pronounced outlier effect in one forecasting item, without which we find that the superforecasting assistant increased accuracy by 41%, compared with 29% for the noisy assistant. We further examine whether LLM forecasting augmentation disproportionately benefits less skilled forecasters, degrades the wisdom-of-the-crowd by reducing prediction diversity, or varies in effectiveness with question difficulty. Our data do not consistently support these hypotheses. Our results suggest that access to a frontier LLM assistant, even a noisy one, can be a helpful decision aid in cognitively demanding tasks compared to a less powerful model that does not provide specific forecasting advice. However, the effects of outliers suggest that further research into the robustness of this pattern is needed.}, number={1}, journal={ACM Transactions on Interactive Intelligent Systems}, author={Schoenegger, Philipp and Park, Peter S. and Karger, Ezra and Trott, Sean and Tetlock, Philip E.}, year={2025}, month=feb, pages={4:1-4:25} }

@inproceedings{Shome_Krishnan_Das_2026, address={San Jose CA USA}, title={Why Johnny Can’t Use Agents: Industry Aspirations vs. User Realities with AI Agents}, ISBN={979-8-4007-2415-2}, url={https://dl.acm.org/doi/10.1145/3786335.3813140}, DOI={10.1145/3786335.3813140}, abstractNote={There is growing imprecision about what “AI agents” are, what they can do, and how effectively they can be used by their intended users. We pose two key research questions: (i) How does the tech industry conceive and market “AI agents”? (ii) What challenges do end-users face when attempting to use commercial AI agents for their advertised uses? We first performed a systematic review of marketed use cases for 102 commercial AI agents, finding that they fall into three umbrella categories: orchestration, creation, and insight. We then evaluated whether end-users could realize these marketed capabilities in practice: we conducted a usability assessment where �� = 31 participants attempted representative tasks for each of these categories on two popular commercial AI agent tools: Operator and Manus. We found that users were generally impressed with these agents but faced significant usability challenges ranging from agent capabilities that were misaligned with user mental models to agents lacking the meta-cognitive abilities necessary for effective collaboration.}, booktitle={Proceedings of the ACM Conference on AI and Agentic Systems}, publisher={ACM}, author={Shome, Pradyumna and Krishnan, Sashreek and Das, Sauvik}, year={2026}, month=may, pages={597–619}, language={en} }

@article{kale_visual_2021,
    title = {Visual {Reasoning} {Strategies} for {Effect} {Size} {Judgments} and {Decisions}},
    volume = {27},
    issn = {1941-0506},
    url = {https://ieeexplore.ieee.org/abstract/document/9222364},
    doi = {10.1109/TVCG.2020.3030335},
    number = {2},
    urldate = {2026-08-19},
    journal = {IEEE Transactions on Visualization and Computer Graphics},
    author = {Kale, Alex and Kay, Matthew and Hullman, Jessica},
    month = feb,
    year = {2021},
    pages = {272--282},
}

@article{padilla_uncertain_2021,
    title = {Uncertain {About} {Uncertainty}: {How} {Qualitative} {Expressions} of {Forecaster} {Confidence} {Impact} {Decision}-{Making} {With} {Uncertainty} {Visualizations}},
    volume = {11},
    issn = {1664-1078},
    shorttitle = {Uncertain {About} {Uncertainty}},
    url = {https://www.frontiersin.org/journals/psychology/articles/10.3389/fpsyg.2020.579267/full},
    doi = {10.3389/fpsyg.2020.579267},
    language = {English},
    urldate = {2026-08-09},
    journal = {Frontiers in Psychology},
    publisher = {Frontiers},
    author = {Padilla, Lace M. K. and Powell, Maia and Kay, Matthew and Hullman, Jessica},
    month = jan,
    year = {2021},
}

@inproceedings{kay_when_2016,
    address = {San Jose California USA},
    title = {When (ish) is {My} {Bus}?: {User}-centered {Visualizations} of {Uncertainty} in {Everyday}, {Mobile} {Predictive} {Systems}},
    isbn = {978-1-4503-3362-7},
    shorttitle = {When (ish) is {My} {Bus}?},
    url = {https://dl.acm.org/doi/10.1145/2858036.2858558},
    doi = {10.1145/2858036.2858558},
    language = {en},
    urldate = {2026-01-12},
    booktitle = {Proceedings of the 2016 {CHI} {Conference} on {Human} {Factors} in {Computing} {Systems}},
    publisher = {ACM},
    author = {Kay, Matthew and Kola, Tara and Hullman, Jessica R. and Munson, Sean A.},
    month = may,
    year = {2016},
    pages = {5092--5103},
}

@misc{Bansal_2024,
      title={Challenges in Human-Agent Communication}, 
      author={Gagan Bansal and Jennifer Wortman Vaughan and Saleema Amershi and Eric Horvitz and Adam Fourney and Hussein Mozannar and Victor Dibia and Daniel S. Weld},
      year={2024},
      eprint={2412.10380},
      archivePrefix={arXiv},
      primaryClass={cs.HC},
      url={https://arxiv.org/abs/2412.10380}, 
}

@article{Sapkota_2026, title={AI Agents vs. Agentic AI: A Conceptual taxonomy, applications and challenges}, volume={126}, ISSN={1566-2535}, url={https://www.sciencedirect.com/science/article/pii/S1566253525006712}, DOI={https://doi.org/10.1016/j.inffus.2025.103599}, abstractNote={Information fusion, in the context of the Generative AI era, must distinguish AI Agents from Agentic AI. This review critically distinguishes between AI Agents and Agentic AI, offering a structured, conceptual taxonomy, application mapping, and analysis of opportunities and challenges to clarify their divergent design philosophies and capabilities. We begin by outlining the search strategy and foundational definitions, characterizing AI Agents as modular systems driven and enabled by LLMs and LIMs for task-specific automation. Generative AI is positioned as a precursor providing the foundation, with AI agents advancing through tool integration, prompt engineering, and reasoning enhancements. We then characterize Agentic AI systems, which, in contrast to AI Agents, represent a paradigm shift marked by multi-agent collaboration, dynamic task decomposition, persistent memory, and coordinated autonomy. Through a chronological evaluation of architectural evolution, operational mechanisms, interaction styles, and autonomy levels, we present a comparative analysis across both AI agents and agentic AI paradigms. Application domains enabled by AI Agents such as customer support, scheduling, and data summarization are then contrasted with Agentic AI deployments in research automation, robotic coordination, and medical decision support. We further examine unique challenges in each paradigm including hallucination, brittleness, emergent behavior, and coordination failure, and propose targeted solutions such as ReAct loops, retrieval-augmented generation (RAG), automation coordination layers, and causal modeling. This work aims to provide a roadmap for developing robust, scalable, and explainable AI-driven systems.}, journal={Information Fusion}, author={Sapkota, Ranjan and Roumeliotis, Konstantinos I. and Karkee, Manoj}, year={2026}, pages={103599} }

@article{Schoenegger_Park_2023,
  title={Large language model prediction capabilities: Evidence from a real-world forecasting tournament},
  author={Schoenegger, Philipp and Park, Peter S},
  journal={arXiv preprint arXiv:2310.13014},
  year={2023}
}

@article{Abolghasemi_Ganbold_Rotaru_2024,
  title={Humans vs. large language models: Judgmental forecasting in an era of advanced AI},
  author={Abolghasemi, Mahdi and Ganbold, Odkhishig and Rotaru, Kristian},
  journal={International Journal of Forecasting},
  volume={41},
  number={2},
  pages={631--648},
  year={2025},
  publisher={Elsevier}
}

@article{Schoenegger_Tuminauskaite_Park_Bastos_Tetlock_2024,
  title={Wisdom of the silicon crowd: LLM ensemble prediction capabilities rival human crowd accuracy},
  author={Schoenegger, Philipp and Tuminauskaite, Indre and Park, Peter S and Bastos, Rafael Valdece Sousa and Tetlock, Philip E},
  journal={Science Advances},
  volume={10},
  number={45},
  pages={eadp1528},
  year={2024},
  publisher={American Association for the Advancement of Science}
}

@article{Halawi_Zhang_Chen_Steinhardt_2024,
  title={Approaching human-level forecasting with language models},
  author={Halawi, Danny and Zhang, Fred and Yueh-Han, Chen and Steinhardt, Jacob},
  journal={Advances in Neural Information Processing Systems},
  volume={37},
  pages={50426--50468},
  year={2024}
}

@article{Zou_2022,
  title={Forecasting future world events with neural networks},
  author={Zou, Andy and Xiao, Tristan and Jia, Ryan and Kwon, Joe and Mazeika, Mantas and Li, Richard and Song, Dawn and Steinhardt, Jacob and Evans, Owain and Hendrycks, Dan},
  journal={Advances in Neural Information Processing Systems},
  volume={35},
  pages={27293--27305},
  year={2022}
}

@inproceedings{jin2021forecastqa,
  title={Forecastqa: A question answering challenge for event forecasting with temporal text data},
  author={Jin, Woojeong and Khanna, Rahul and Kim, Suji and Lee, Dong-Ho and Morstatter, Fred and Galstyan, Aram and Ren, Xiang},
  booktitle={Proceedings of the 59th Annual Meeting of the Association for Computational Linguistics and the 11th International Joint Conference on Natural Language Processing (Volume 1: Long Papers)},
  pages={4636--4650},
  year={2021}
}

@book{Tetlock_Gardner_2015,
  title={Superforecasting: The art and science of prediction},
  author={Tetlock, Philip Eyrikson and Gardner, Dan and Gardner, Dan},
  volume={221},
  year={2015},
  publisher={Crown Publishers New York}
}

@book{Tetlock_2005,
  title={Expert political judgment: How good is it? How can we know?},
  author={Tetlock, Philip E},
  year={2009},
  publisher={Princeton University Press}
}

@book{Armstrong_2001,
  title={Principles of forecasting: a handbook for researchers and practitioners},
  author={Armstrong, Jon Scott and others},
  volume={30},
  year={2001},
  publisher={Springer}
}

@inproceedings{yan2024autocast,
  title={Autocast++: Enhancing world event prediction with zero-shot ranking-based context retrieval},
  author={Yan, Qi and Seraj, Raihan and He, Jiawei and Meng, Lili and Sylvain, Tristan},
  booktitle={International Conference on Learning Representations},
  volume={2024},
  pages={34523--34539},
  year={2024}
}

@article{pratt_2024,
  title={Can language models use forecasting strategies?},
  author={Pratt, Sarah and Blumberg, Seth and Carolino, Pietro Kreitlon and Morris, Meredith Ringel},
  journal={arXiv preprint arXiv:2406.04446},
  year={2024}
}

@misc{Metaculus_AI_forecasting_2026, url={https://www.metaculus.com/notebooks/43363/ai-forecasting-in-2026/}, abstractNote={This post is a synthesis of 11 Metaculus analyses run between Oct 2024 and May 2026, seeking to summarize everything we know about AI forecasting and how to do it well. Conclusions from other papers, blogs, and benchmarks are also discussed in order to give a comprehensive overview of the field.}, language={en} }

@misc{metaculus2026apichanges,
  author       = {Unnaamed},
  title        = {Changes to the {M}etaculus {API}},
  howpublished = {\url{https://www.metaculus.com/notebooks/42554/changes-to-the-metaculus-api/}},
  year         = {2026},
  month        = {mar},
  note         = {Metaculus notebook; accessed 2026-09-07}
}

@misc{kalshi2026,
  author       = {{Kalshi Inc.}},
  title        = {Kalshi: A Regulated Exchange and Prediction Market},
  howpublished = {\url{https://kalshi.com}},
  year         = {2026},
  note         = {Accessed 2026-09-07}
}

@misc{polymarket2026,
  author       = {{Polymarket}},
  title        = {Polymarket: The World's Largest Prediction Market},
  howpublished = {\url{https://polymarket.com}},
  year         = {2026},
  note         = {Accessed 2026-09-07}
}

@article{Paleka_Goel_Geiping_Tramèr_2025, title={Pitfalls in Evaluating Language Model Forecasters}, url={http://arxiv.org/abs/2506.00723}, DOI={10.48550/arXiv.2506.00723}, abstractNote={Large language models (LLMs) have recently been applied to forecasting tasks, with some works claiming these systems match or exceed human performance. In this paper, we argue that, as a community, we should be careful about such conclusions as evaluating LLM forecasters presents unique challenges. We identify two broad categories of issues: (1) difficulty in trusting evaluation results due to many forms of temporal leakage, and (2) difficulty in extrapolating from evaluation performance to real-world forecasting. Through systematic analysis and concrete examples from prior work, we demonstrate how evaluation flaws can raise concerns about current and future performance claims. We argue that more rigorous evaluation methodologies are needed to confidently assess the forecasting abilities of LLMs.}, note={arXiv:2506.00723 [cs.LG]}, number={arXiv:2506.00723}, publisher={arXiv}, author={Paleka, Daniel and Goel, Shashwat and Geiping, Jonas and Tramèr, Florian}, year={2025}, month=may, language={en} }

@article{Turtel_Wilczewski_Franklin_Skothiem_2026, title={Future-as-Label: Scalable Supervision from Real-World Outcomes}, url={http://arxiv.org/abs/2601.06336}, DOI={10.48550/arXiv.2601.06336}, abstractNote={Time creates free supervision: forecasts about real-world events resolve to verifiable outcomes. The passage of time provides labels that require no annotation. To exploit this structure, we extend reinforcement learning with verifiable rewards to real-world prediction over time. We train language models to make probabilistic forecasts from causally masked information, using proper scoring rules as the reward function once events resolve. Learning is driven entirely by realized outcomes, enabling scalable outcome-based supervision in open-world prediction. On real-world forecasting benchmarks, Qwen3-32B trained using Foresight Learning improves Brier score by 27% and halves calibration error relative to its pretrained baseline, and outperforms Qwen3-235B on both constructed future-event prediction tasks and the Metaculus benchmark despite a 7x parameter disadvantage.}, note={arXiv:2601.06336 [cs.LG]}, number={arXiv:2601.06336}, publisher={arXiv}, author={Turtel, Benjamin and Wilczewski, Paul and Franklin, Danny and Skothiem, Kris}, year={2026}, month=jan }

@article{Turtel_Franklin_Schoenegger_2025, title={LLMs Can Teach Themselves to Better Predict the Future}, url={http://arxiv.org/abs/2502.05253}, DOI={10.48550/arXiv.2502.05253}, abstractNote={We present an outcome-driven fine-tuning framework that enhances the forecasting capabilities of large language models (LLMs) without relying on human-curated reasoning samples. Our method leverages model self-play to generate pairs of diverse reasoning trajectories and probabilistic forecasts for a set of diverse questions that resolve after the models’ knowledge cutoff date. We then rank pairs of these reasoning traces by their distance to the actual outcomes before fine-tuning the model via Direct Preference Optimization (DPO). On a separate test set, our approach increases prediction accuracy of Phi-4 14B and DeepSeek-R1 14B by between 7--10% over a base model and a DPO fine-tuned control model with randomized labels, bringing them on par with forecasting capabilities of much larger frontier models like GPT-4o.}, note={arXiv:2502.05253 [cs.CL]}, number={arXiv:2502.05253}, publisher={arXiv}, author={Turtel, Benjamin and Franklin, Danny and Schoenegger, Philipp}, year={2025}, month=feb }

@article{Alur_2025, title={AIA Forecaster: Technical Report}, url={http://arxiv.org/abs/2511.07678}, DOI={10.48550/arXiv.2511.07678}, abstractNote={This technical report describes the AIA Forecaster, a Large Language Model (LLM)-based system for judgmental forecasting using unstructured data. The AIA Forecaster approach combines three core elements: agentic search over high-quality news sources, a supervisor agent that reconciles disparate forecasts for the same event, and a set of statistical calibration techniques to counter behavioral biases in large language models. On the ForecastBench benchmark (Karger et al., 2024), the AIA Forecaster achieves performance equal to human superforecasters, surpassing prior LLM baselines. In addition to reporting on ForecastBench, we also introduce a more challenging forecasting benchmark sourced from liquid prediction markets. While the AIA Forecaster underperforms market consensus on this benchmark, an ensemble combining AIA Forecaster with market consensus outperforms consensus alone, demonstrating that our forecaster provides additive information. Our work establishes a new state of the art in AI forecasting and provides practical, transferable recommendations for future research. To the best of our knowledge, this is the first work that verifiably achieves expert-level forecasting at scale.}, note={arXiv:2511.07678 [cs.AI]}, number={arXiv:2511.07678}, publisher={arXiv}, author={Alur, Rohan and Stadie, Bradly C. and Kang, Daniel and Chen, Ryan and McManus, Matt and Rickert, Michael and Lee, Tyler and Federici, Michael and Zhu, Richard and Fogerty, Dennis and Williamson, Hayley and Lozinski, Nina and Linsky, Aaron and Sekhon, Jasjeet S.}, year={2025}, month=nov }

@misc{ForecastBench, url={https://www.forecastbench.org/}, abstractNote={ForecastBench is a dynamic, contamination-free benchmark of LLM forecasting accuracy with human comparison groups, serving as a valuable proxy for general intelligence.}, journal={ForecastBench}, language={en-US} }

@article{Velardo_Elliott_2021, title={Co-Interviewing in Qualitative Social Research: Prospects, Merits and Considerations}, volume={20}, ISSN={1609-4069}, url={https://doi.org/10.1177/16094069211054920}, DOI={10.1177/16094069211054920}, abstractNote={In this paper, we seek to open a dialogue about the approach of co-interviewing, which, to our surprise, has not received much attention in the realm of qualitative social research. The co-interviewing approach stands apart from ‘multiple’ interviewing, in which two or more researchers are tasked with conducting interviews in a single research study. Instead, this approach involves two researchers actively taking part in the same interview. In a qualitative grounded theory study that sought to explore doctoral students’ emotional well-being, we adopted the approach of co-interviewing each of our participants. This process involved us sharing the responsibilities of asking questions, probing, note taking and making observations. Our experience has led us to critically examine this unique approach to interviewing, and here, in this paper, we offer insights about its potential to enhance the generation of data and the research process. In doing so, we draw on literature, whilst reflecting on key concepts including power, reflexivity and well-being, by considering the positions of participants and researchers alike.}, journal={International Journal of Qualitative Methods}, publisher={SAGE Publications Inc}, author={Velardo, Stefania and Elliott, Sam}, year={2021}, month=jan, pages={16094069211054920}, language={EN} }

@inproceedings{Cai2024,
author = {Cai, Mandi and Kay, Matthew},
title = {Watching the Election Sausage Get Made: How Data Journalists Visualize the Vote Counting Process in U.S. Elections},
year = {2024},
isbn = {9798400703300},
publisher = {Association for Computing Machinery},
address = {New York, NY, USA},
url = {https://doi.org/10.1145/3613904.3642329},
doi = {10.1145/3613904.3642329},
booktitle = {Proceedings of the 2024 CHI Conference on Human Factors in Computing Systems},
articleno = {391},
numpages = {19},
location = {Honolulu, HI, USA},
series = {CHI '24}
}

@inproceedings{10.1145/3290605.3300803,
author = {Guo, Shunan and Du, Fan and Malik, Sana and Koh, Eunyee and Kim, Sungchul and Liu, Zhicheng and Kim, Donghyun and Zha, Hongyuan and Cao, Nan},
title = {Visualizing Uncertainty and Alternatives in Event Sequence Predictions},
year = {2019},
isbn = {9781450359702},
publisher = {Association for Computing Machinery},
address = {New York, NY, USA},
url = {https://doi.org/10.1145/3290605.3300803},
doi = {10.1145/3290605.3300803},
booktitle = {Proceedings of the 2019 CHI Conference on Human Factors in Computing Systems},
pages = {1–12},
numpages = {12},
location = {Glasgow, Scotland Uk},
series = {CHI '19}
}

@inproceedings{10.1145/3313831.3376866,
author = {Sun, Dong and Feng, Zezheng and Chen, Yuanzhe and Wang, Yong and Zeng, Jia and Yuan, Mingxuan and Pong, Ting-Chuen and Qu, Huamin},
title = {DFSeer: A Visual Analytics Approach to Facilitate Model Selection for Demand Forecasting},
year = {2020},
isbn = {9781450367080},
publisher = {Association for Computing Machinery},
address = {New York, NY, USA},
url = {https://doi.org/10.1145/3313831.3376866},
doi = {10.1145/3313831.3376866},
booktitle = {Proceedings of the 2020 CHI Conference on Human Factors in Computing Systems},
pages = {1–13},
numpages = {13},
location = {Honolulu, HI, USA},
series = {CHI '20}
}

@inproceedings{10.1145/3491102.3502010,
author = {Koval, Morgane and Jansen, Yvonne},
title = {Do You See What You Mean? Using Predictive Visualizations to Reduce Optimism in Duration Estimates},
year = {2022},
isbn = {9781450391573},
publisher = {Association for Computing Machinery},
address = {New York, NY, USA},
url = {https://doi.org/10.1145/3491102.3502010},
doi = {10.1145/3491102.3502010},
booktitle = {Proceedings of the 2022 CHI Conference on Human Factors in Computing Systems},
articleno = {30},
numpages = {19},
location = {New Orleans, LA, USA},
series = {CHI '22}
}

@inproceedings{10.1145/3613904.3642371,
author = {Yang, Fumeng and Mortenson, Chloe Rose and Nisbet, Erik and Diakopoulos, Nicholas and Kay, Matthew},
title = {In Dice We Trust: Uncertainty Displays for Maintaining Trust in Election Forecasts Over Time},
year = {2024},
isbn = {9798400703300},
publisher = {Association for Computing Machinery},
address = {New York, NY, USA},
url = {https://doi.org/10.1145/3613904.3642371},
doi = {10.1145/3613904.3642371},
booktitle = {Proceedings of the 2024 CHI Conference on Human Factors in Computing Systems},
articleno = {389},
numpages = {24},
location = {Honolulu, HI, USA},
series = {CHI '24}
}

@inproceedings{10.1145/2858036.2858150,
author = {Badam, Sriram Karthik and Zhao, Jieqiong and Sen, Shivalik and Elmqvist, Niklas and Ebert, David},
title = {TimeFork: Interactive Prediction of Time Series},
year = {2016},
isbn = {9781450333627},
publisher = {Association for Computing Machinery},
address = {New York, NY, USA},
url = {https://doi.org/10.1145/2858036.2858150},
doi = {10.1145/2858036.2858150},
booktitle = {Proceedings of the 2016 CHI Conference on Human Factors in Computing Systems},
pages = {5409–5420},
numpages = {12},
location = {San Jose, California, USA},
series = {CHI '16}
}

@misc{Giangiulio_2026, title={Looking at just the odds isn’t enough. How traders gain an edge on prediction markets}, url={https://www.cnbc.com/2026/08/01/traders-go-full-time-on-prediction-markets-using-ai-bots-and-antennas.html}, abstractNote={Gaining an edge on prediction markets requires speed and skills beyond just reading the odds on platforms like Kalshi and Polymarket.}, journal={CNBC}, author={Giangiulio, Ananya Chetia, Davis}, year={2026}, month=aug, language={en} }

@misc{spring2026survey, url={https://www.metaculus.com/notebooks/45382/bot-survey-spring-2026/}, author={Wilson, Ben}, language={en-US}, year={2026} }
